# CORRESPONDENCE

## To the Editors of 'The Observatory'

*Early solar photographs by G. Roster (April 1893)*

A great effort has been made in the last twenty years to obtain a better reconstruction of the Sunspot Number using historical observations that cover the last four centuries[1-4]. Moreover, spatial information about sunspots in the past is being recovered by several teams. The information includes Butterfly diagrams[5-6], sunspot catalogues[7], and even sunspot group tilt records[8].

The use of antique solar photographs to increase our databases or simply to check well-known data[9] is of great interest. However, there are relatively few photographs of the Sun compared to drawings of the solar disc[10]. Therefore, we want to highlight three solar photographs taken by Giorgio Roster in 1893. Roster (1843-1894) was a doctor, chemist, and photographer. He was interested in the usefulness of the photography in sciences and in works on microphotography and telephotography[11]. The library of the *Museo di storia della scienza* of Florence preserves part of Roster's legacy in the section *Carte e raccolta fotografica Roster,* in which we want to highlight the title *Telefotografie Roster V: Macchie solari; teleobiettivo Roster, 1892-1893* (https://bibdig.museogalileo.it/Teca/Viewer?an=964650). This title contains three solar photographs taken from Florence with a Roster tele-objective. The Roster tele-objective permitted to modify the enlargement due to the possibility of separating or bringing closer the positive and negative elements of the tele-objective. The comparison among the Roster photography and the Greenwich Photoheliographic Results (GPR) permits to infer the position of the solar north. The main characteristics of the three photographs are: (*i*) Date: 21st April 1893. Time: 5:15 p.m. Magnification: 61x. (*ii*) Date: 25th April 1893. Time: 4:40 p.m. Magnification: 68x. (*iii*) Date: 30th April 1893. Time: 17:00 p.m. Magnification: 71x.

It is important to note that there are some defects in the objective that could be confused with sunspots, although these imperfections are easily identifiable comparing the three photographs. In any case, sunspot groups are visible just coinciding with its heliographic coordinates provided by the GPR. In addition, the photographs are not correctly oriented. The solar north will point to the top of the photographs if they are rotated -81º, 99º, and 99º respectively. Finally, it may be interesting to note that, although there are drawings of the solar disc in the collection of the Kalocsa Observatory[12] for the days 21 and 30, there is no such thing for day 25. Thus, Roster's photographs fill that gap in the graphical information of the Sun.

We appreciate the support of EU, Junta de Extremadura, and Ministry of Economy and Competitiveness (consortium IMDROFLOOD, Research Group Grant GR15137, IB16127, and AYA2014-57556-P).

Yours faithfully,

FERNANDO DOMÍNGUEZ-CASTRO

Instituto Pirenaico de Ecología

Consejo Superior de Investigaciones Científicas (IPE-CSIC)

Avda. Montañana, 1005.

E-50059 Zaragoza, Spain

JOSÉ M. VAQUERO

Departamento de Física

Centro Universitario de Mérida, Universidad de Extremadura

Avda. Santa Teresa de Jornet, 38

E-06800 Mérida, Badajoz, Spain

2017 October 12